\documentstyle[aps,prb,graphicx,floats]{revtex}
\begin{document}
\title{The temperature--flow renormalization group 
and the competition between superconductivity and ferromagnetism}
\author{Carsten Honerkamp$^{1}$ and Manfred Salmhofer$^{2}$}

\address{$^{1}$ Theoretische Physik,
ETH-H\"onggerberg, CH-8093 Z\"urich, Switzerland \\
$^{2}$ Theoretische Physik, Universit\"at Leipzig, 
D-04109 Leipzig, Germany} 
\date{May 10, 2001}
\maketitle
\begin{abstract} 
We derive a differential equation for the one--particle--irreducible vertex functions of interacting fermions as a function of the temperature.
Formally, these equations correspond to a Wilsonian 
renormalization group (RG) scheme which uses the temperature as an explicit
scale parameter. Our novel method allows us to analyze the competition between
superconducting and various magnetic Fermi surface instabilities
in the one--loop approximation. 
In particular this includes ferromagnetic fluctuations, which are 
difficult to treat
on an equal footing in conventional Wilsonian momentum space techniques. 
Applying the scheme to the
two-dimensional $t$-$t'$ Hubbard model we investigate the RG flow of
the interactions at the van Hove filling 
with varying next-nearest neighbor hopping $t'$. Starting at
$t'=0$ we describe
the evolution of the flow to strong coupling from an antiferromagnetic nesting 
regime over a $d$-wave regime at moderate $t'$ to a
ferromagnetic region at larger absolute values of $t'$.  Upon increasing the
particle density in the latter regime the ferromagnetic tendencies are cut off
and the leading instability occurs in the triplet superconducting pairing channel.
\end{abstract}

\section{Introduction}
Wilsonian momentum shell renormalization group (RG) schemes 
have become an important tool for the
weak-coupling analysis of interacting electron systems
because they provide a means to sum competing classes of diagrams in perturbation theory 
in a consistent manner.
Moreover, with the Wilsonian RG schemes one can systematically approach
 potential singularities, 
which would cause problems in a direct perturbation expansion. 
Successes of RG methods in interacting electron systems are for 
example clear arguments for the stability
of the Landau-Fermi liquid in two and more spatial dimensions for short range repulsive
interactions and non-nested Fermi surfaces unless a Kohn-Luttinger instability 
intervenes at lowest temperatures \cite{msbook,shankar,froehlich}, or the
classification of ground states in quasi-one-dimensional systems
\cite{solyom,emery,voit}. 
Moreover, numerous one-loop RG calculations have been used recently to analyze
the interplay between antiferromagnetic (AF) and $d$-wave 
superconducting (SC) pairing tendencies in the 
two-dimensional (2D) Hubbard model
\cite{dzialoshinski,schulz,lederer,gonzales,zanchi,alvarez,furukawa,halboth,honerkamp,honerkamp2}.
On the other hand, the Hubbard model was originally introduced 
to describe ferromagnetism (FM) of itinerant electrons (see, e.g., Refs. \onlinecite{yoshida,fazekas}). The existence of spin--polarized ground states
has been proven in  
one--dimensional \cite{fazekas,tasaki} and in special higher--dimensional
\cite{mielke} models and also established in the limit of infinitely
many dimensions \cite{ulmke,vollhardt}. In the $t-t'$ Hubbard model on the 2D
square lattice at weak to moderate $U$,  Hartree--Fock results
\cite{linhirsch}, $T$-matrix approximation together with Quantum Montecarlo 
simulations \cite{hlubina1,hlubina2} and  a generalized random phase
approximation approach including local particle-particle correlations\cite{fleck}
 point towards a FM regime for larger
absolute values of $t'$ around the van Hove filling where the Fermi surface
(FS) contains the saddle points at $(\pm \pi,0)$ and $(0,\pm
\pi)$. Recent parquet calculations\cite{irkhin} indicated similar tendencies. 
However, momentum--shell RG methods have up to now not produced any evidence confirming these findings.

In this paper we introduce a novel RG scheme, which we call the 
temperature--flow RG. Its distinguishing feature is that the 
temperature itself plays the role of the flow parameter. 
Indeed, it has been observed in many studies that varying a 
cutoff scale is in many ways similar to varying the temperature,
but here we provide an exact RG flow equation for the 
one--particle--irreducible correlation
functions, parameterized by the temperature. 
Using this method, we show that in the one--loop approximation,
one finds a ferromagnetic regime in the $t-t'$ Hubbard model. 
More precisely, the new flow is similar to previous ones
for small $t'$ and near to half--filling, 
but for larger $|t'|$, a ferromagnetic
regime appears. The boundary  of the
FM phase is roughly at the same $t'_c$ where 
it would be in the Hartree--Fock approximation,
but with our RG calculation we find a {\em quantum critical point}  separating 
the FM regime from a $d$-wave superconducting regime instead of a first order
transition to an AF phase as in Hartree--Fock.
We also find a triplet superconducting state when the filling 
is increased above the van Hove value. 

In the following, we discuss the reasons why this FM phase
was not seen in previous RG calculations. The main point is
that when the flow parameter is given by an infrared cutoff,
the particle--hole excitations with small total momentum are 
artificially suppressed above the scale set by 
the temperature, but in many interesting cases one cannot 
run the flow down to these low scales because the one--loop RG flow 
of the interactions leads to strong coupling and even diverges at a nonzero scale.

In RG flows to strong coupling, 
the main interest is to determine the nature of the strongly coupled
state. Very often there are several candidates for the latter 
and it is natural to assume that the strong coupling state with the 
highest energy gain will be realized.
In the RG treatment this state is typically associated with the
dominant channel at the energy scale at which the RG flow diverges, e.g.  determined 
through an analysis of the corresponding susceptibilities.
However, one--loop flows become unreliable once the coupling constants
get larger than a value which depends, among other factors, on the Fermi surface. 
Thus it can happen that the one--loop approximation 
breaks down before processes involving
particle-hole fluctuations at small wave-vectors, which 
could drive ferromagnetism or screen longer--ranged interactions,
can intervene. Because of its inequivalent treatment
of FM fluctuations on the one side and AF and SC tendencies on the other side, 
this type of momentum shell RG with infrared cutoff does presumably not
represent, at least when used too straightforwardly,
an appropriate method to answer the question
which of the channels prevails\cite{fnkopietz}.  

Thus, although momentum shell schemes apply, 
and can even be controlled rigorously, as long as the running coupling 
constants do not get large, it is interesting to investigate and 
compare alternative schemes which are not based on strict infrared cutoffs 
and therefore capable of taking into account low--energy fluctuations at
all wave-vectors in the same way. 

We now specifically discuss the Stoner problem.
Consider an electron system with a density of states
$\rho$ per spin orientation and
a repulsive onsite interaction $U>0$.
The RPA spin susceptibility is  
\begin{equation} \chi_s (\omega,\vec{q}) = \frac{ \chi^0_s (\omega, 
\vec{q})}{ 1 -  U \chi^0_s (\omega, \vec{q}) } \, . \label{rpasusnew}
\end{equation}  
where $\chi^0_s (\omega, \vec{q})$ is the bare particle--hole bubble
\begin{equation} 
\chi^0_s (\omega, \vec{q}) = -
\int \frac{d^Dk}{(2 \pi)^D}
\, \frac{n_F(\epsilon (\vec{k}+\vec{q}) )-n_F(\epsilon (\vec{k}) )}{
i\omega +\epsilon
(\vec{k}+\vec{q})-\epsilon (\vec{k})} 
\end{equation}
with $n_F$ the Fermi distribution. Sending $\omega \to 0$
and afterwards $\vec{q} \to 0$, we get
\begin{equation}
 \chi^0_s (\omega=0, \vec{q} \to 0) = 
 - \int \frac{d^Dk}{(2 \pi)^D}
\frac{d}{d\epsilon} n_F (\epsilon (\vec{k}) ) 
=
\int dE \, \rho(E) \, \delta_T (E)
\label{chiph0}
\end{equation}
where $\delta_T (E) = - {d \over dE} n_F(E) = (4T\cosh^2(E/2T))^{-1}$
becomes the delta function as $T \to 0$. If the density of states
$\rho$ is continuous, we get, at $T=0$, $ \chi_s (\omega=0, \vec{q} \to 0) =
\rho(0) / (1-\rho(0) U)$. The pole in this expression leads to the 
simple Stoner criterion
\begin{equation}\label{stonercrit}
U \rho(0) \ge 1
\end{equation}  
for a ferromagnetic ground state.

Of course, it is well known that the Stoner criterion
overestimates the tendency towards ferromagnetism, and one should include
correlation effects. One of them is that in (\ref{stonercrit}),
$U$ is expected to get replaced by 
an effective Stoner coupling $U_{\mathrm{Stoner}}$.
This was first discussed in the so-called $T$-matrix approximation by
Kanamori\cite{kanamori}. On the other hand, the $T$--matrix 
approximation also sums only a subclass of diagrams of one--loop type.
Thus the natural question arises whether by RG methods one could perform
a consistent resummation of all one--loop contributions,
to treat the ferromagnetic tendencies and the screening
together with other possibly competing tendencies.  

Let us attempt to apply a one-loop momentum-shell scheme to the Stoner problem.
The divergence of the uniform spin susceptibility is equivalent to a pole in
the RPA perturbation series for the forward scattering amplitude
$a^a(\vec{k},\vec{k}',\vec{q} \to 0)$ which is obtained from the 
antisymmetric irreducible two-particle scattering vertex $\Gamma^{(4)}$ via
\begin{eqnarray} 
a^a (\vec{k}, \vec{k}' )  &=& \lim_{\vec{q} \to 0} \Gamma^{(4)} \left[  (\vec{k}+\vec{q},s,
\vec{k}',-s) \to (\vec{k},-s,
\vec{k}'+\vec{q},s)  \right] \label{aspin} \\ &=&
\lim_{\vec{q} \to 0} \left\{ \Gamma^{(4)} \left[  (\vec{k}+\vec{q},s,
\vec{k}',s) \to (\vec{k},s,
\vec{k}'+\vec{q},s)  \right] - \Gamma^{(4)} \left[  (\vec{k}+\vec{q},s,
\vec{k}',-s) \to (\vec{k},s,
\vec{k}'+\vec{q},-s)  \right] \right\} \, . 
\nonumber \end{eqnarray}
Since the
$\vec{q} \to 0$ particle-hole loop is the only diagram occuring in the
RPA calculation of $a^a (\vec{k}, \vec{k}' )$ and also in the RPA 
calculation leading to the Stoner criterion, 
we can limit the RG analysis to this channel, in
complete analogy to Cooper or SDW (spin density wave) 
instabilities, where the ladder summation
of the zero-total-momentum particle-particle diagram or the particle-hole
diagram with the appropriate momentum transfer, respectively, are written as
solutions of the one-loop RG equation for the corresponding channel
only (as indicated above, we do a study taking into account all 
contributions below). The RG equation is then the differential equation
\begin{equation}\label{quaDE}
{\partial \over \partial \Lambda} U_\Lambda = - R_\Lambda U_\Lambda^2,
\end{equation}  
where $R_\Lambda (\omega,\vec{q})$ is given by the integral for $\chi_s^0$,
but with the integrand containing an additional factor 
${\partial \over \partial \Lambda} [h(\epsilon(\vec{k}+\vec{q})/\Lambda)
h(\epsilon(\vec{k})/\Lambda)]$, with $h(x)$ the cutoff function,
say, $h(x) =0$ for $|x| \le 1$. 
In the limit of interest for FM,
\begin{equation}\label{scaled}
R_\Lambda (\omega=0,\vec{q} \to 0) =
{\partial \over \partial \Lambda} F_\Lambda (T) \, ,
\qquad
F_\Lambda (T) =
\int dE \, \rho(E) \delta_T (E)  
\, h(E/\Lambda)^2 \, ,
\end{equation}  
and (\ref{quaDE}) integrates to 
\begin{equation}
U_\Lambda = \frac{U_{\Lambda_0}}%
{1- U_{\Lambda_0} (F_\Lambda(T) - F_{\Lambda_0}(T))}
\end{equation} 
If $\Lambda_0$ is larger than the bandwidth, $ F_{\Lambda_0}(T)=0$.
For $\Lambda \to 0$, $F_\Lambda(T) \to \chi_s^0 (0,0)$, 
given by (\ref{chiph0}),
so we recover exactly the RPA result, and thus the Stoner criterion. 

However, a closer look at (\ref{scaled}) reveals that $F_\Lambda(T)$
is negligibly small for all scales $\Lambda$ above $O(T)$
because $\delta_T(E) \le e^{-|E|/T}$ and the cutoff function $h$
restricts to $|E| \ge \Lambda$. Thus the
contribution to the RPA is built up only at the very lowest scales
of the flow: small-$\vec{q}$ particle-hole
fluctuations, such as ferromagnetic tendencies, 
occur only at low RG scales $\Lambda \approx T$\cite{fn1,fnalvarez}. 
For the derivation of the RPA as a solution to an 
approximate RG, this is not a problem. But our goal is to
understand the interplay of different instabilities, and it 
can happen that another instability driven by terms that 
contribute to the flow at higher scales gives rise to a flow to strong
coupling at scales $\Lambda > T$. 
Then the small-$\vec{q}$ contributions would remain
unintegrated because, as discussed above, 
we have to stop the one-loop scheme once the couplings
cease to be small. This appears to be particularly unjustified if the density
of states (DOS) at the FS is much larger than the DOS at higher band energies, 
e.g. in cases where the FS is close to a van Hove singularity.

One can argue that this problem is not only due to the 
one--loop approximation but that it also has to do
with restricting to $\vec{q}=0$. If $|\vec{q}|$ is of order $
\Lambda/v_F$, the infrared cutoff does not suppress the particle--hole
susceptibility any more. Thus one problem is the nonuniformity in the
buildup of the function of $\vec{q}$, and one might think that averaging
over $\vec{q}$ in a region of size $O(\Lambda/v_F)$ may help.

However, in this paper we attack the problem directly:
as we shall show, in the temperature--flow RG scheme, the small
$\vec{q}$ fluctuations are properly taken into account directly, 
without further approximations like the averaging mentioned above. 

In the next section, we set up the exact RG equation for the 
temperature--flow, and describe the one--loop approximation.
After considering simple examples we 
apply the scheme to the two-dimensional $t$-$t'$ repulsive Hubbard
model on the square lattice. This model is interesting
because it supports a variety of fluctuations which can become singular 
at low temperatures and an unbiased treatment is highly desirable.
On one hand it shows strong antiferromagnetic fluctuations due to
partial nesting of the Fermi surface at finite wavevectors and omnipresent
superconducting instabilities at low temperatures are 
enhanced by the anisotropic scattering. On the other hand the 2D Hubbard
model is also a good candidate for a ferromagnetic ground
state even at weak coupling because of the large density of 
states at the Fermi surface for certain particle densities. 
 
\section{Formalism}
Here we derive differential equations which determine the
evolution of the 1PI (one-particle-irreducible) vertex
functions with varying temperature. A perturbative  RG scheme describing the
change of the 1PI functions with variation of a general
parameter-dependent quadratic part of the free action, $S_0^{(2)}$,
was  described in Refs. \onlinecite{salmhofer} and \onlinecite{honerkamp}.  
In these applications the continuous 
change in $S_0^{(2)}$ corresponded to integrating out modes with band
energy at the running RG scale $\Lambda$. Here we show 
how the same RG equations can be used to describe the variation of the 1PI
vertex functions with changing temperature. In this case the scale parameter
can be directly taken as the temperature of the model system. 

Consider a fermionic system on a 2D lattice at temperature $T$, with action
\begin{eqnarray} S &=&  T \sum_n \sum_s \int \frac{d^2k}{(2 \pi)^2} \, \bar{\psi}_s
(\vec{k}, i \omega_n) \, ( i \omega_n - \epsilon_{\vec{k}} ) \psi_s (\vec{k}, i
\omega_n) \nonumber \\ && + \frac{1}{2} \, T^3  
\sum_{n_1,n_2,n_3} \sum_{s,s'} \int
\frac{d^2k_1}{(2 \pi)^2} \frac{d^2k_2}{(2 \pi)^2} \frac{d^2k_3}{(2 \pi)^2} V
(\vec{k}_1, \vec{k}_2, \vec{k}_3) \bar{\psi}_s
(\vec{k_3}, i \omega_{n_3}) \bar{\psi}_{s'} (\vec{k_4}, i \omega_{n_4})
\psi_{s'} (\vec{k_2}, i \omega_{n_2}) \psi_{s} (\vec{k_1}, i \omega_{n_1}) \nonumber \\ && +
\sum_n \sum_s \int \frac{d^2k}{(2 \pi)^2} \, \left[ 
\bar{\psi}_s (\vec{k}, i \omega_n)  \xi _s (\vec{k}, i \omega_n)  + 
\psi_s (\vec{k}, i \omega_n)  \bar{\xi} _s (\vec{k}, i \omega_n) \right] 
 \, . \label{psiaction}
\end{eqnarray}
Here $\bar{\psi}$ and $\psi$ are anticommuting Grassmann fields and the Matsubara frequencies $\omega_n=(2n+1)\pi T$ are summed over all integer $n$.
We have assumed spin-rotation-invariant and frequency-independent
interactions. Further we have added source fields $\bar{\xi}$ and $\xi$. Obviously both the quadratic and the quartic part of the action
depend on the temperature $T$. 

We now recall the case where only the quadratic part of the action
depends on an additional continuous 
parameter $\Lambda$. Formally we go from  
$S$ to a $\Lambda$--dependent action $S_\Lambda$ by replacing
\begin{equation}
T \, \left[  i \omega - \epsilon (\vec{k}) \right] \quad \longrightarrow \quad
Q_\Lambda (i \omega, \vec{k} ) .
\end{equation}  
One can then derive  RG differential equations describing the variation 
of the 1PI vertex functions with $\Lambda$. These equations are 
obtained from the $\Lambda$-dependent Legendre transform
\begin{equation} \Gamma_\Lambda (\phi , \bar{\phi}) = W_\Lambda  (\xi, \bar{\xi}) - \sum_n \sum_s \int 
\frac{d^2k}{(2\pi)^2} \,  \left[ 
\bar{\phi}_s (\vec{k}, i \omega_n)  \xi _s (\vec{k}, i \omega_n)  + 
\phi_s (\vec{k}, i \omega_n)  \bar{\xi} _s (\vec{k}, i \omega_n) \right]  \, , 
\label{legtraf} \end{equation}
of the generating functional of the connected non-amputated 
$m$--point Green functions
$W^{(m)}$,
\begin{equation} 
e^{-W_\Lambda (\xi,\bar{\xi})} = \int D\psi D\bar{\psi} \; e^{-S_\Lambda  (\psi, \bar{\psi},
\xi,\bar{\xi})} \, . \label{genfuncW} \end{equation}
\begin{figure} 
\begin{center} 
\includegraphics[width=.6\textwidth]{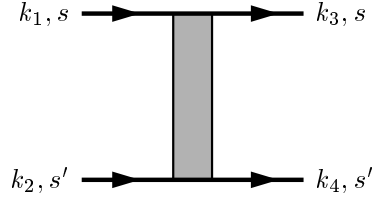}
\end{center}
\caption{The coupling function $V(\vec{k}_1,\vec{k}_2,\vec{k}_3)$. $\vec{k}_4$
is fixed by wavevector conservation on the lattice.
 }
\label{v4}
\end{figure}

\begin{figure} 
\begin{center} 
\includegraphics[width=.4\textwidth]{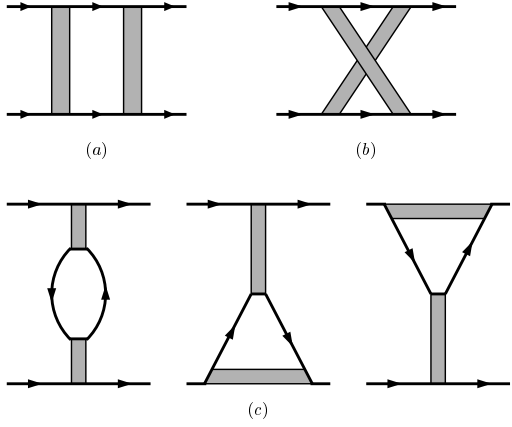}
\end{center}
\caption{The particle-particle and particle-hole diagrams contributing to the
one-loop RG equation.
 }
\label{rgdia}
\end{figure}

Here we abbreviate the notation and define a combined index $p=(i \omega_n,
\vec{k},s)$, moreover $\int dp = \sum_{n,s} \int \frac{d^2k}{(2\pi)^2}$.
The RG differential equations describing  the evolution of the selfenergy $\Sigma_\Lambda (p)$ and
coupling function $V_\Lambda (p_1,p_2,p_3)$ with $p_4=p_1+p_2-p_3$ (from which the fully
antisymmetric four-point vertex can be 
reconstructed\cite{salmhofer}, see also Fig. \ref{v4}) with variation of that parameter read
\begin{equation} \frac{d}{d\Lambda} \Sigma_\Lambda (p) = \int dp' S_\Lambda(p') \,  \left[ V_\Lambda ( p,p',p') 
- 2 V_\Lambda ( p,p',p) \right] \label{sigmadot} \end{equation}
and 
\begin{equation} \frac{d}{d\Lambda} V_\Lambda (p_1,p_2,p_3) =    {\cal T}_{PP,\Lambda} + {\cal T}^d_{PH,\Lambda} +
{\cal T}^{cr}_{PH,\Lambda} \label{vdot} \end{equation}
with
\begin{eqnarray}
\lefteqn{  {\cal T}_{PP,\Lambda} (p_1,p_2;p_3,p_4) = }\nonumber\\ 
&& - \int dp \, 
V_\Lambda ( p_1,p_2,p ) \, L(p,-p+p_1+p_2) \, V_\Lambda (p,-p+p_1+p_2 ,p_3) \label{PPdia}  
\\
 \lefteqn{ {\cal T}^d_{PH,\Lambda} (p_1,p_2;p_3,p_4) =}  \nonumber \\
&&- \int dp\,
\biggl[ -2 V_\Lambda ( p_1,p,p_3 ) \, L(p,p+p_1-p_3) \, V_\Lambda (p+p_1-p_3,p_2,p) 
\nonumber 
 \\ && \qquad +
V_\Lambda (p_1,p,p+p_1-p_3) \, L(p,p+p_1-p_3) \, V_\Lambda (p+p_1-p_3,
p_2,p)
 \nonumber  \\ && \qquad +
V_\Lambda ( p_1,p,p_3) \, L(p,p+p_1-p_3) \, V_\Lambda (p_2,p+p_1-p_3,p)
\biggr]\label{PHddia} 
\\
\lefteqn{  {\cal T}^{cr}_{PH,\Lambda}(p_1,p_2;p_3,p_4) =} \nonumber \\ 
&& - \int dp \, 
V_\Lambda (p_1,p+p_2-p_3,p)  \, L(p,p+p_2-p_3) \, V_\Lambda (p,p_2,p_3 )\label{PHcrdia}
\end{eqnarray}
In these equations, 
\begin{equation} \label{e13}
L(p,p') = S_\Lambda (p) W^{(2)}_{\Lambda} (p') + W^{(2)}_{\Lambda} (p) S_\Lambda (p')  \, \end{equation}
with the so-called single-scale propagator
\begin{equation}
S_\Lambda (p) = - W^{(2)}_{\Lambda} (p) \left[ \frac{d}{d\Lambda} Q_\Lambda
(p) \right] \, W^{(2)}_{s} (p) \, .
\end{equation}  
The one-loop diagrams corresponding to the terms (\ref{PPdia}), (\ref{PHddia}) and
(\ref{PHcrdia})  are shown in Fig. \ref{rgdia}.

In the typical momentum-shell RG, the varying parameter $\Lambda$ is an energy
scale which separates high and low energy modes. The strategy is to
integrate out the high energy modes first. In this case the scale parameter
only affects the quadratic part $Q_\Lambda (p)$, which is multiplied with an
appropriate cutoff function. 
Here, for the reasons discussed in the introduction,  we want to treat the
temperature as varying parameter. 
Our reasoning is as follows:
at high temperatures, where $\pi T$ is larger than the bandwidth and the 
interaction energies, perturbation theory converges, and 
moreover the corrections to the  selfenergy and 
four--point function are of order $1/T$, hence small.
Thus at high temperature, the vertex functions are essentially identical
to the terms in the action. Then we track the
renormalization of the vertex functions when the temperature is lowered.  Of
course this idea is implicit in most of the well-known scaling approaches
\cite{solyom}. On a technical level however this strategy has usually been
cast into some cutoff-variation procedure with similar results as the
more elaborate modern Wilsonian schemes.

In order to apply the RG scheme for the 1PI vertex functions
with $T$ as a flow parameter
we first have to perform a transformation which shifts all temperature 
dependence to the quadratic part of the action.

\subsection{New fields} The $T^3$-factor in the interaction part can be 
removed by transforming the action to the new fermionic fields given by
\begin{equation}  \bar{\eta}_s  (\vec{k}, n) = T^{3/4} \bar{\psi}_s
(\vec{k}, i \omega_n) \, , \quad \eta_s  (\vec{k}, n) = T^{3/4} \psi_s
(\vec{k},  i \omega_n) \, . \label{newfields} \end{equation}
The source fields are chosen to transform according to 
\begin{equation}  \bar{\zeta}_s  (\vec{k}, n) = T^{-3/4} \bar{\xi}_s
(\vec{k},  i \omega_n) \, , \quad \zeta_s  (\vec{k}, n) = T^{-3/4} \xi_s
(\vec{k}, i \omega_n) \, . \label{newsources} \end{equation}
The action then reads
\begin{eqnarray} S & =&  T^{-1/2} \sum_n \sum_s \int \frac{d^2k}{(2 \pi)^2} \, \bar{\eta}_s
(\vec{k}, n) \, ( i\omega_n - \epsilon_{\vec{k}} ) \eta (\vec{k}, n)  \nonumber \\ && 
+ \frac{1}{2}  \sum_{{n_1},{n_2},{n_3} } \sum_{s,s'} \int
\frac{d^2k_1}{(2 \pi)^2} \frac{d^2k_2}{(2 \pi)^2} \frac{d^2k_3}{(2 \pi)^2} V
(\vec{k}_1, \vec{k}_2, \vec{k}_3) \bar{\eta}_s
(\vec{k_3}, {n_3}) \bar{\eta}_{s'} (\vec{k_4}, {n_4})
\eta_{s'} (\vec{k_2}, {n_2}) \eta_{s} (\vec{k_1}, {n_1}) \nonumber \\ && +
\sum_{n} \sum_s \int \frac{d^2k}{(2 \pi)^2} \, \left[ 
\bar{\eta}_s (\vec{k}, n)  \zeta _s (\vec{k}, n)  + 
\eta_s (\vec{k}, n)  \bar{\zeta} _s (\vec{k}, n) \right] 
 \, . \label{etaaction}
\end{eqnarray}
Only the quadratic part depends on the temperature. 
The connected $m$--point correlation functions $W^{(m)}_{\psi}$ and
$W^{(m)}_{\eta}$ are related  as follows (the change in the integration
measure drops out):
\begin{eqnarray}  
W^{(m)}_{\psi} (\vec{k}_1, i\omega_{n_1} , \dots \vec{k}_m, i\omega_{n_m} ) &=& \langle \psi_{s_1} (\vec{k}_1,i\omega_{n_1})  \dots \bar{\psi}_{s_m}
(\vec{k}_m, i\omega_{n_m})  \rangle \nonumber \\ &=& T^{-3m/4}  \langle \eta_{s1} (\vec{k}_1,n_1)  \dots \bar{\eta}_{s_m}
(\vec{k}_m,n_m)  \rangle \nonumber \\ &=& T^{-3m/4}  W^{(m)}_{\eta}
(\vec{k}_1, {n_1} , \dots \vec{k}_m, {n_m} ) \label{wmpointrel} 
\, . \end{eqnarray} 
Using the action in terms of the new fields we can now apply the RG formalism
from Ref. \onlinecite{salmhofer}. This yields the flow of vertex functions
$\Gamma^{(m)}_{\eta} (\vec{k}_1, {n_1} , \dots \vec{k}_m, {n_m} ) $ when we
change the temperature. 
Since
they are the expansion coefficients of $\Gamma_s$, which is a functional 
of of fields conjugated to the $\zeta$ source fields, the
$\Gamma^{(m)}_{\eta}$ are related to the vertex functions of the $\psi$ fields via
\begin{equation}  \Gamma^{(m)}_{\psi} (\vec{k}_1, i\omega_{n_1} , \dots \vec{k}_m, i\omega_{n_m} ) = T^{3m/4}  \Gamma^{(m)}_{\eta}
(\vec{k}_1, {n_1} , \dots \vec{k}_m, {n_m} ) \label{mpointrel} 
\, , \end{equation} in particular
\begin{equation}
\Gamma^{(2)}_{\psi} (\vec{k}_1, i\omega_{n_1} , \vec{k}_2, i\omega_{n_2} ) =
T^{3/2}  \Gamma^{(2)}_{\eta} (\vec{k}_1, n_1 ,  \vec{k}_2, {n_2} )
\end{equation}  
and
\begin{equation}
\Gamma^{(4)}_{\psi} (\vec{k}_1, i\omega_{n_1} , \dots \vec{k}_4, i\omega_{n_4} ) =
T^{3}  \Gamma^{(4)}_{\eta} (\vec{k}_1, n_1 , \dots  \vec{k}_4, {n_4} ) \,
.
\end{equation}  
Let us consider an example. If the interaction term at temperature
$T$ is taken to be
\begin{equation}  \frac{1}{2} \, T^3 \sum_{n_1,n_2,n_3} \sum_{s,s'} \int
\frac{d^2k_1}{(2 \pi)^2} \frac{d^2k_2}{(2 \pi)^2} \frac{d^2k_3}{(2 \pi)^2} V_{T}
(\vec{k}_1, \vec{k}_2, \vec{k}_3) \bar{\psi}_s
(\vec{k_3}, i \omega_{n_3}) \bar{\psi}_{s'} (\vec{k_4}, i \omega_{n_4})
\psi_{s'} (\vec{k_2}, i \omega_{n_2}) \psi_{s} (\vec{k_1}, i \omega_{n_1}) \, ,
\label{T0int} \end{equation}
the antisymmetric four-point vertex in the $\psi$-fields
reads\cite{salmhofer}, to first order in the interaction,
\begin{equation}  \Gamma^{(4)}_{\psi, T} (\vec{k}_1,s, i \omega_{n_1};\vec{k}_2,s', i \omega_{n_2};\vec{k}_3,s, i \omega_{n_3};\vec{k}_4,s', i \omega_{n_4})= T^3 \left[
V_{T}(\vec{k}_1, \vec{k}_2, \vec{k}_3) - \delta_{s,s'} V_{T}(\vec{k}_2, \vec{k}_1,
\vec{k}_3) \right] \, . \end{equation}
The four-point vertex for the $\eta$-fields is simply 
\begin{equation}
\Gamma^{(4)}_{\eta,
T} (\vec{k}_1,s, {n_1};\vec{k}_2,s', {n_2};\vec{k}_3,s, {n_3};\vec{k}_4,s', {n_4}) 
 = T^{-3} \,  \Gamma^{(4)}_{\psi, T} (\vec{k}_1,s, i \omega_{n_1};\vec{k}_2,s', i
\omega_{n_2};\vec{k}_3,s, i \omega_{n_3};\vec{k}_4,s', i \omega_{n_4}) \, ,
\end{equation}  
hence of order $T^0$, and it contains the coupling function
$V_{T}$. Thus the RG equations for the four-point-vertex in terms of the
$\eta$-fields exactly yield the flow of the coupling function, and we can use
the relations (\ref{mpointrel}) to express the results in terms of the
original vertex functions.

In this paper, like in Refs. \onlinecite{honerkamp} and
\onlinecite{honerkamp2}, we will neglect selfenergy corrections to the flow and furthermore truncate the
infinite hierarchy of RG equations for the $m$-point vertex functions by
setting all 1PI vertex function higher than fourth order equal to zero. These
approximations restrict the validity of the results to the weakly coupled and weakly
correlated regime. We also drop the frequency dependence of the four-point vertex.

\subsection{The single-scale propagator} The quadratic part of the action
expressed in the $\eta$-fields reads
$ Q_{\eta,T} (\vec{k}, n )= T^{-1/2} ( i\omega_n - \epsilon_{\vec{k}} )$.
In absence of selfenergy corrections, we obtain
\begin{equation}\label{W}
W_{\eta,T}^{(2)}(\vec{k}, n )= \frac{T^{1/2} }{ i\omega_n - \epsilon_{\vec{k}} } \,
.
\end{equation}
Therefore the single-scale propagator is given by
\begin{equation}\label{S}  S_T (\vec{k}, n ) = - W_{\eta,T}^{(2)}(\vec{k}, n ) \, \left[ \frac{d}{dT} Q_{\eta,T}(\vec{k},
n ) \right] \, W_{\eta,T}^{(2)}(\vec{k}, n )  = -\frac{T^{-1/2} }{2} \, \frac{i
\omega_n + \epsilon (\vec{k}) }{\left[ i\omega_n -\epsilon
(\vec{k})\right]^2} \, . \label{siscaprop} \end{equation} 

\subsection{One-loop diagrams} \label{IIc}
Next, consider the product of two fermion propagators $L(p,p')$
defined in (\ref{e13}), which appears in the RG equation described in
\onlinecite{salmhofer}. The index $p$ contains the frequency $\omega$ and wavevector $\vec{k}$,
at band energy $\epsilon $ associated to one propagator 
and $p'$ the frequency $\omega'$, wavevector
$\vec{k}'$ at band energy $\epsilon'$ associated to the other one. 
Inserting (\ref{W}) and (\ref{S}) into (\ref{e13}), we have 
\begin{equation}
L(p,p') =
- \frac{1}{2} 
\left[ 
\frac{i\omega + \epsilon }{(i\omega - \epsilon )^2} 
\frac{1}{i \omega' - \epsilon'} 
+ \frac{1}{i \omega - \epsilon} 
\frac{i \omega' + \epsilon' }{(i\omega' - \epsilon' )^2} 
\right]
= \frac{\omega \omega' + \epsilon \epsilon'}%
{(i\omega - \epsilon)^2 (i \omega' - \epsilon')^2}
=
\frac{d}{dT} \left[ T   
\frac{1}{i\omega - \epsilon} \frac{1}{i\omega' - \epsilon'} \right] \, .
\end{equation}
The terms on the right hand side of the flow equation are of the form
\begin{equation}
\int dp\; L(p,p') \Phi (\vec{k},s,\vec{k}',s')
=
\int \frac{d^2 k}{(2\pi)^2} \, \sum_{s,s'}
\frac{d}{dT} \left[ T \sum_{n}   
\frac{1}{i\omega_n - \epsilon(\vec{k})} \frac{1}{i\omega_{n'} - \epsilon(\vec{k'})} \right] \, 
\Phi (\vec{k},s,\vec{k}',s'),
\end{equation}
i.e.\ they are temperature derivatives of bubble diagrams. 
Thus,  in order to calculate the temperature flow of
the coupling function, we can  think in terms of 
the usual one-loop diagrams of the perturbation expansion in the original
$\psi$-fields and simply take the temperature derivative of the particle-hole
and particle-particle bubbles.

\section{Simple examples}
Here we apply the scheme to two simple examples and show that it yields the
results expected from direct perturbation expansions. In particular we describe
how the temperature-flow scheme allows us to describe Cooper and ferromagnetic 
instabilities in an analogous fashion. 

\subsection{Cooper instability}
Here we show that for  a small attractive electron-electron interaction at
temperature $T_0$ the formalism 
leads to a Cooper instability at lower temperature $T_c$. We assume that at $T = T_0$ the action is given by the BCS model,
\begin{eqnarray} S_{T_0} &=& T_0 \sum_n \int\frac{d^2k}{(2 \pi)^2}  \,  \bar{\psi}  (\vec{k}, i \omega_n)
\, ( i \omega_n - v_F (k -k_F))   \, \psi  (\vec{k}, i \omega_n) \nonumber \\
&& + \, \frac{1}{2} \,  
V^{\mathrm{Cooper}}_{T_0} \, T_0^3 
\sum_{n_1,n_2,n_3} \int \frac{d^2k}{(2\pi)^2}
\int \frac{d^2k'}{(2\pi)^2} \, \bar{\psi}_s
(\vec{k}', i \omega_{n_3}) \bar{\psi}_{-s} (-\vec{k}', i \omega_{n_4})
\psi_{-s} (-\vec{k}, i \omega_{n_2}) \psi_{s} (\vec{k}, i \omega_{n_1}) \, 
. \label{bcsaction} \end{eqnarray} 
Here, $V^{\mathrm{Cooper}}_{T_0}<0$ is  the attractive interaction between electrons with opposite
wavevectors and the bandwidth $W$ is taken such that $v_F k_F -W > 0$. We assume a linear dispersion. 
We are interested in the low-temperature behavior of the
four-point vertex with zero 
incoming total momentum and frequency, 
\[ \Gamma_{\psi,T}^{(4)}\left[  (\vec{k},s,-\vec{k},-s) \to (\vec{k}',s,-\vec{k}',-s)\right] =
T^3 V_T (\vec{k},-\vec{k}, \vec{k}') \, . \]
Focusing on the pairing channel, we neglect all particle-hole
contributions. Therefore we are 
left with the temperature derivative of the particle-particle bubble and 
$V_T (\vec{k},-\vec{k}, \vec{k}') = V_T^{\mathrm{Cooper}}$ remains independent of $\vec{k}$ and
$\vec{k}'$.
The RG equation for $V^{\mathrm{Cooper}}_{T}$ involves the particle--particle
bubble $\Pi(\omega,\vec{q})$ with $\omega=0$ and pair momentum
$\vec{q}=0$. Using the results from Sec.\ \ref{IIc} it reads
\begin{equation} \frac{d}{dT} V^{\mathrm{Cooper}}_{T}= -V^{\mathrm{Cooper}}_T\ ^2 \, \frac{d}{dT} \Pi (0,0) =-
V^{\mathrm{Cooper}}_T\ ^2 \,  \frac{d}{dT}
\int_{-W}^W d\epsilon \, \rho(\epsilon) \,  
\frac{1-2n_F (\epsilon )}{2\epsilon}  \, . \label{rgcooper}
\end{equation}
We can immediately integrate it, similarly to (\ref{scaled}), to get
\begin{equation}\label{intecoop}
- \frac{1}{V^{\mathrm{Cooper}}_T} \Big\vert_{T_0}^T 
=
- \int_{-W}^W d\epsilon \, \rho(\epsilon) 
\frac{1-2 n_F (\epsilon)}{2 \epsilon}\Big\vert_{T_0}^T \; .
\end{equation}
Assuming that the bandwidth $W$ is much larger than the temperature
and that the density of states is a constant $\rho_0$, 
the integral is logarithmic and 
\begin{equation} 
V^{\mathrm{Cooper}}_T = \frac{V^{\mathrm{Cooper}}_{T_0} }{1- \rho_0 V^{\mathrm{Cooper}}_{T_0} \log (T/T_0) } \,  . \label{coopersol}
\end{equation}
Thus, starting with an attractive interaction $V^{\mathrm{Cooper}}_{T_0} <0 $ at a higher temperature
$T_0 \ll W$ we obtain a pole in the effective $s$-wave  Cooper pair scattering
at
\begin{equation}
T_c = T_0 \, \exp\left[ - \frac{1}{\rho_0 \left| V^{\mathrm{Cooper}}_{T_0}\right|} \right] \,
. \label{coopertc} 
\end{equation}
This result is analogous to the ladder summation from straightforward
perturbation theory when we replace $T_0$ with the Debye frequency and
identify the initial interaction $V^{\mathrm{Cooper}}_{T_0}$ at this 
temperature with the bare
attractive electron-electron interaction $V^{\mathrm{Cooper}}_0$, 
e.g. mediated through phonon-exchange.
Although these identifications cannot be justified directly within a
microscopic model with a specific Hamiltonian, they appear to be highly
plausible for an effective description of the system. 

In fact one can obtain the BCS relation 
\begin{equation} T_c =1.14 W \exp \left[ - \frac{1}{\rho_0 |V^{\mathrm{Cooper}}_{T_0}|} \right]
\label{bcstcrel} \end{equation}
as well if one assumes that $T_0\gg W$. In that case, inserting 
$T_0$ in the right hand side of (\ref{intecoop}) gives zero
because at high temperatures $n_F \approx 1/2$. Thus 
we recover the linearized BCS gap equation\cite{fw} at $T_c$, 
\begin{equation}  \frac{1}{\rho_0 |V^{\mathrm{Cooper}}_{T_0}|} =  \int_{-W}^{W} d\epsilon \,
\frac{1-2n_F (\epsilon )}{2\epsilon} \, , \label{bcsgapeqtc} \end{equation}
as a condition for a divergence of $V^{\mathrm{Cooper}}_{T}$ at $T=T_c$. From solving (\ref{bcsgapeqtc}),
one obtains (\ref{bcstcrel}). 

\subsection{Stoner instability}
Next we consider the instability towards ferromagnetism in a Hubbard-type system with
repulsive onsite interaction $U_{\mathrm{eff}}$. We show that for a logarithmically
divergent density of states within an effective bandwidth
$W_{\mathrm{eff}}$,
\begin{equation} \rho (\epsilon ) = \bar{\rho} \, \log \frac{ W_{\mathrm{eff}}
}{|\epsilon |} \label{divdos} \, ,  \end{equation}
an approximation to the RG equation reproduces the Stoner instability
obtained by straightforward perturbation theory. In the latter treatment one
sums up the particle-hole bubbles for momentum transfer $\vec{q} \to 0$ with the
intermediate particles having band energies within 
$\pm W_{\mathrm{eff}}$. The van Hove singularity in the density of states causes 
a breakdown of the paramagnetic state at temperatures below
\begin{equation} 
T_c \sim W_{\mathrm{eff}} \, \exp \left[ - \frac{1}{U_{\mathrm{eff}} \bar{\rho}}
\right] \, .\label{rpasumtc} \end{equation}
Analogous to the diagram summation we restrict the RG analysis 
to the crossed particle-hole channel with small momentum transfer. 
The Stoner instability should then appear as a
divergence of the forward scattering amplitude $a^a(\vec{k},\vec{k}',\vec{q}
\to 0)$, which is in our RG language given by
\[ \lim_{\vec{q} \to 0} \Gamma_{\psi,T}^{(4)} \left[ (\vec{k},s,\vec{k}'+\vec{q} ,-s)
\to ( \vec{k}',s,\vec{k}+\vec{q},-s) \right] = \lim_{\vec{q} \to 0}
T^3 V_T (\vec{k},\vec{k}'+\vec{q}, \vec{k}') \, . \] 
Next we will assume an isotropic system and write
\[ \lim_{\vec{q} \to 0} V_T (\vec{k},\vec{k}'+\vec{q}, \vec{k}') = V^{\mathrm{FM}}_T \, . \]
Using Eq. (\ref{chiph0}) with (\ref{divdos}), the RG equation for this coupling function reads
\begin{equation}\label{Tdif}
\frac{d}{dT} V^{\mathrm{FM}}_{T} = - V^{\mathrm{FM}}_T\ ^2 \, \frac{d}{dT}
\chi_{s} (0,\vec{q} \to 0) = -V^{\mathrm{FM}}_T\ ^2 \, \frac{d}{dT}
\int_{-W}^W d\epsilon \,  \rho (\epsilon) \, \frac{d}{d\epsilon} n_F (\epsilon ) \, .
\end{equation}
Note the similarity of, but also the difference between, (\ref{Tdif}) and (\ref{scaled}) --
both equations have the same form, but there is no cutoff function 
in (\ref{Tdif}). Integration of (\ref{Tdif}) gives
\begin{equation}
\frac{1}{ V^{\mathrm{FM}}_{T_0}} - \frac{1}{ V^{\mathrm{FM}}_{T}} = F(T) - F(T_0), 
\end{equation}
with
\begin{equation}
\qquad F(T) = \chi^0_s (0,0) = \int dE \, \rho(E) \delta_T (E)
\end{equation}
Because $\delta_T (E) \propto T^{-1}$, $F(T_0)$ vanishes as $T_0 \to \infty$,
hence we recover for large $T_0$ the RPA result
\begin{equation}
V^{\mathrm{FM}}_T = \frac{ V^{\mathrm{FM}}_T}{1- V^{\mathrm{FM}}_T F(T)}
\end{equation}
In the case of the density of states (\ref{divdos}), we can also 
use integration by parts to get, for $T_0 \ll W$,
 \[ \frac{d}{dT} V^{\mathrm{FM}}_{T}= - V^{\mathrm{FM}}_T\ ^2 \bar{\rho}  \, \int_{-W}^W d \epsilon \,
\frac{1}{\epsilon} \, \frac{d}{dT}\,  n_F (\epsilon ) =  V^{\mathrm{FM}}_T\ ^2 \bar{\rho}  \, \frac{1}{T} \, \int_{-W}^W d \epsilon \,
\frac{d}{d\epsilon}\,  n_F (\epsilon ) \, .  \]
The resulting  differential equation is solved by
\begin{equation} 
V^{\mathrm{FM}}_T = \frac{V^{\mathrm{FM}}_{T_0} }{1 +   V^{\mathrm{FM}}_{T_0} \bar{\rho} \log (T/T_0) } \,  . \label{stonersol}
\end{equation}
Analogous to the Cooper instability studied above, the initial interaction
$V_{T_0} $ at higher temperature $T_0 \ll W$ , in this case repulsive,
leads to a pole at
\begin{equation}
T_c = T_0 \, \exp\left[ - \frac{1}{\bar{\rho} V^{\mathrm{FM}}_{T_0}} \right] \,
, \label{stonertc} 
\end{equation}
signaling the instability of the paramagnetic state with respect to
spontaneous polarization. Again this result is in close analogy with the RPA
summation when we replace the initial temperature $T_0$ with the effective bandwidth
$W_{\mathrm{eff}}$. Of course this approximation to the RG equation 
can also lead to pole for a
non-divergent density of states if the initial interaction is sufficiently
strong. In this case, upon integrating from $T_0 \gg W$ down towards 
$T=0$, we encounter a pole if the simple Stoner criterion (\ref{stonercrit})
is fulfilled. 

Thus, in our RG treatment, the simple Stoner criterion can be derived
as a consequence of an approximation to the one--loop RG equation 
where only the crossed
particle--hole terms are kept. As we shall see in the next section,
the full one--loop flow is more complicated, and one can thereby see
the limitations of the naive Stoner criterion, and instances where it
becomes invalid because of superconducting and antiferromagnetic 
tendencies.  

\section{Application to the 2D Hubbard model}
Here we apply the RG scheme to the 2D $t$-$t'$ Hubbard model. 
In our setup with the temperature as a flow parameter, this again means that
we assume that at some higher temperature $T_0$ the single-particle
Green's function of the system is adequately described by
\begin{equation}
G_0 (i \omega , \vec{k} ) = \frac{1}{i \omega - \epsilon (\vec{k}) } \, ,
\end{equation}
with the band dispersion $\epsilon (\vec{k}) =-2t \left[ \cos k_x + \cos k_y
\right] -4t' \cos k_x \cos k_y -\mu$ including nearest and next-nearest
neighbor hoppings $t$ and $t'$, respectively. 
The scattering vertex is given by a local repulsion
\begin{equation}  V_{T_0} (\vec{k}_1,\vec{k}_2 , \vec{k}_3 ) = U \, .
\label{iniv} \end{equation} 
Then by following the flow of the scattering vertex with lowering the
temperature, we obtain information on the possible low-temperature phases of
the system. In particular we analyze which classes of coupling functions and
which susceptibilities become important at low temperatures.
For a large parameter range, we observe a flow to strong coupling. 
This means that at sufficiently low temperatures some
components of the coupling function $V_T (\vec{k}_1,\vec{k}_2 , \vec{k}_3 ) $
reach values larger than the bandwidth $\approx 8t$. 
The approximations made, i.e. the restriction to one-loop equations and the
neglect of selfenergy corrections, fail when the couplings get too large. 
Therefore we stop the flow when the largest coupling
exceeds a high value larger than the bandwidth, 
e.g. $V _{T, \mathrm{max}}=18t$. 
This defines a {\em characteristic temperature} $T_c$ 
of the flow to strong coupling. As it is well known, in two spatial dimensions breaking of
continuous symmetries is impossible at $T>0$. Thus the flow to
strong coupling which we observe for a large part of the parameter range
should either be interpreted as an indication of an ordered ground state of
the corresponding type or as a tendency towards ordering at $T \sim T_c$ for the
case when an additional coupling term in the third spatial direction is included. 

\begin{figure} 
\begin{center} 
\includegraphics[width=.4\textwidth]{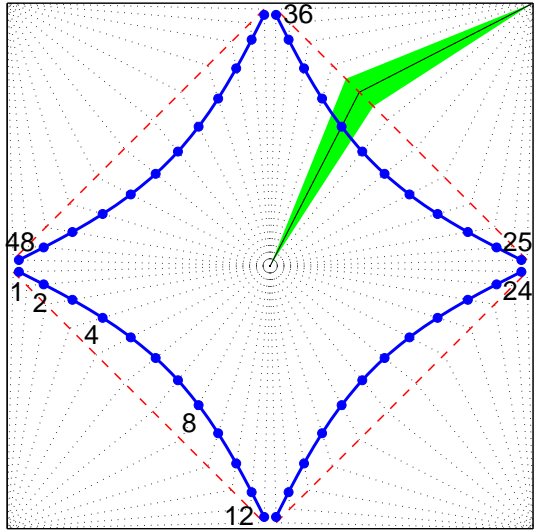}
\end{center}
\caption{The Brillouin zone, Fermi (solid line) and umklapp (dashed line) surface and the lines in the
patch lines for $N= 32$. The circles denote the wave vectors for 
which the four-point vertex function is calculated. For all wave vectors
inside the same 
patch (shaded region) around the lines, $V_\Lambda (\vec{k}_1,\vec{k}_2,\vec{k}_3)$ is approximated by a
constant. 
 }
\label{setup}
\end{figure}

For the numerical integration of these coupled equations we apply
a phase space discretization \cite{zanchi,halboth,honerkamp,honerkamp2}.  
The 2D Brillouin zone (BZ)  is divided up into $N$ elongated patches centered around $N$
lines.  Each line connects
the origin with one of the $(\pm \pi ,\pm \pi)$-points with two straight
pieces and a kink at the Umklapp surface (see Fig. \ref{setup}).

Next we discretize the coupling function. We approximate $V_{T} (\vec{k}_1,\vec{k}_2,\vec{k}_3)$ by a constant for all wave
vectors in the same patches and calculate the RG flow for the
subset of interaction vertices with one wavevector representative for each
patch. We take these wave vectors as the
crossing points of the $N$ lines with the Fermi surface (FS). 
The phase space space integrations
are performed as sums over the patches and integrations over the radial 
direction along 3 or 5 lines inside each patch. Most calculations were using 
 48-patch systems. Calculations with 96 patches gave equivalent results.  
A typical Fermi surface with $N=48$ points is shown in Fig. \ref{setup}.

The motivation for using this patch scheme is that the Fermion propagators
are largest on the Fermi surface and decay away from it. Thus one can expect
that by tracking the coupling functions for the wavevectors 
close to the Fermi surface the leading flow will be described accurately. 

Together with the flow of the interactions we calculate several static
susceptibilities by coupling external fields of appropriate form to the
electrons. During the flow these external couplings are renormalized through
one-loop vertex corrections, as described in Ref. \onlinecite{honerkamp}. In this
paper, we concentrate on the $d_{x^2-y^2}$- and $p$-wave pairing
susceptibilities, 
the AF susceptibility $\chi_s (\vec{q}= (\pi, \pi))$ and the FM susceptibility
$\chi_s (\vec{q} \to 0)$.

\section{Results for the 2D Hubbard model at the van Hove filling}
Here we describe the results for the RG flow of the coupling function 
for the 2D Hubbard model with initial interaction
$U=3t$ and varying value for the next-nearest neighbor hopping $t'$.
The chemical potential $\mu$ is fixed\cite{fn2} at the van Hove value $\mu = 4t'$ such the FS
always contains the saddle points at $(\pm \pi,0)$ and $(0,\pm
\pi)$.  We start
the RG scheme at temperature $T_0 = 4t$ and integrate the coupled equations
with decreasing temperature. We stop the flow when the largest coupling
exceeds a high value larger than the bandwidth, e.g. $V_{T, \mathrm{max.}}
=18t$, which defines the characteristic temperature for the flow to strong
coupling, $T_c$.
 
\begin{figure} 
\begin{center} 
\includegraphics[width=.5\textwidth]{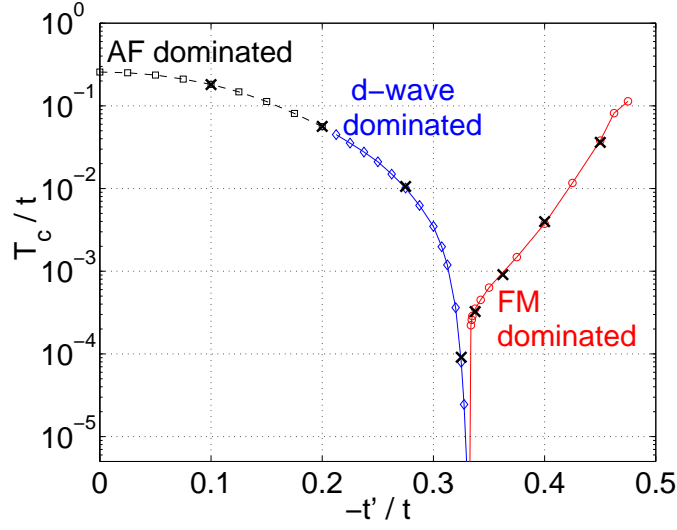}
\end{center}
\caption{Characteristic temperature $T_c$ for the flow to strong coupling versus next
nearest neighbor hopping amplitude $t'$ from a $48$-patch calculation (the crosses 
show data for 96 patches). 
The chemical potential is fixed at
the van Hove value $\mu =4t'$. $T_c$ is defined as temperature where the
couplings reach values larger than $18t$. For small $|t'|$, the instability is 
dominated by AF tendencies, for $t'< -0.2t$ the $d$-wave tendencies prevail,
and for $t'< -0.33t$ the ferromagnetic fluctuations grow strongest. The
criterion for the distinction between these regimes is taken as the
derivative of the susceptibilities with respect to temperature at the scale
where the couplings become larger than $10t$.}
\label{tctpvh}
\end{figure}

\begin{figure} 
\begin{center} 
\includegraphics[width=.6\textwidth]{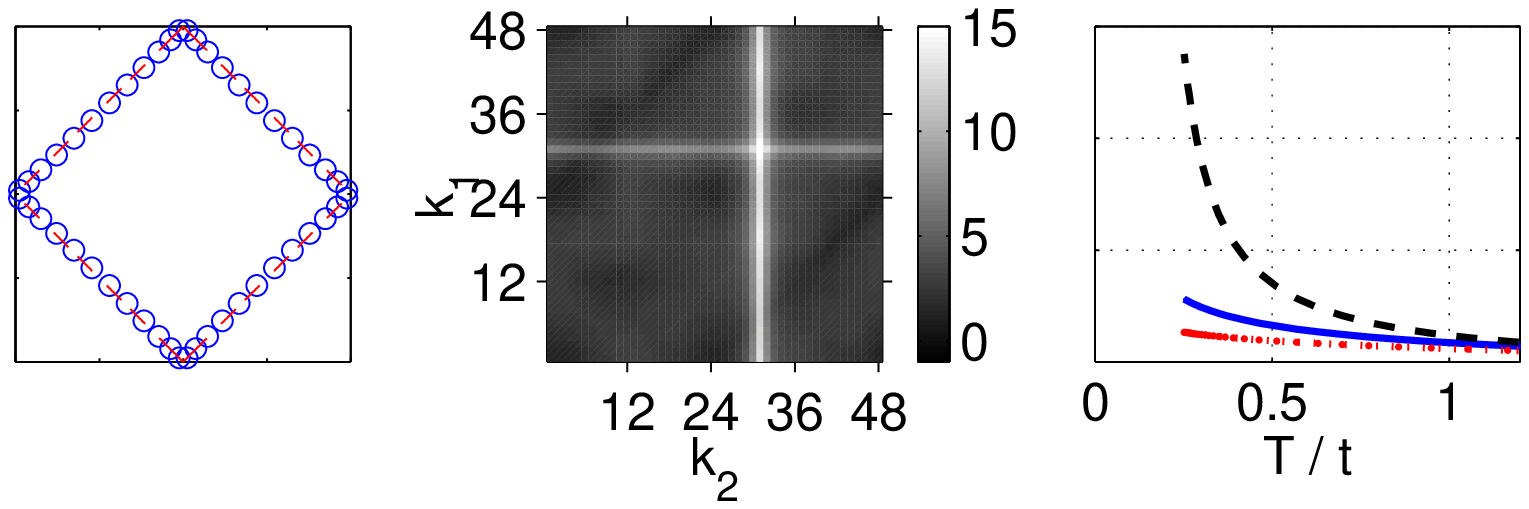}
\end{center}
\caption{Flow to strong coupling for $t'=-0.025t$ and $\mu=-0.1t$. The left
plot shows the 48 points on the Fermi surface (points 1, 12, 13, 24 etc. are
closest to the saddle points). The middle plot shows the
coupling function $V_T(k_1,k_2,k_3)$ at low $T$ as a function of the two
incoming wavevectors labeled by $k_1$ and $k_2$ moving around the FS. $k_3$
is fixed at point 6 in the Brillouin zone diagonal. The colorbar denotes the value of the
coupling function in units of $t$. The right plot shows the flow of the AF
susceptibility (dashed line), $d$-wave susceptibility (solid line) and FM
susceptibility (dotted line) as a function of the temperature $T$.}
\label{vp10}
\end{figure}

The overall behavior of the flow to strong coupling is shown in
Fig. \ref{tctpvh}. Focusing on AF, $d$-wave and FM susceptibilities, 
we observe three distinct parameter regions. The first, the AF regime, 
occurs closer to half band filling, 
for smaller absolute values of $t'>-0.2t$. Here the flow
to strong coupling takes place at relatively high temperatures and the AF
susceptibility is growing most strongly (see Fig. \ref{vp10} for the Fermi
surface and the flow of the susceptibilities). The dominant scattering processes at
low $T$ are given by AF processes between FS parts connected by wavevectors
$\approx (\pi , \pi)$.  These can be seen in the middle plot of Fig. \ref{vp10} 
as bright features corresponding to strong repulsive couplings on the 
line $\vec{k_2} - \vec{k_3} \approx (\pi , \pi)$.

\begin{figure} 
\begin{center} 
\includegraphics[width=.6\textwidth]{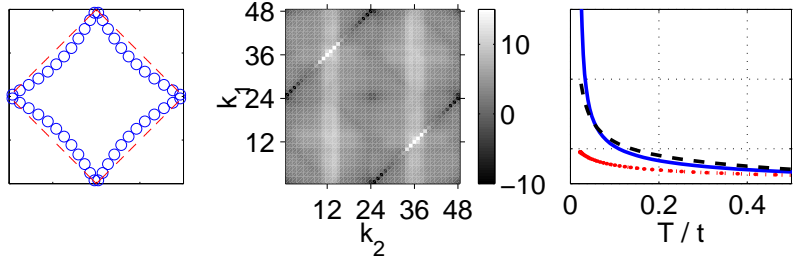}
\end{center}
\caption{Flow to strong coupling for $t'=-0.25t$ and $\mu=-t$. The left
plot shows the 48 points on the Fermi surface (points 1, 12, 13, 24 etc. are
closest to the saddle points). The middle plot shows the
coupling function $V_T(k_1,k_2,k_3)$ at low $T$ as a function of the two
incoming wavevectors labeled by $k_1$ and $k_2$ moving around the FS. $k_3$
is fixed at point 1 close to a saddle point. The colorbar denotes the value of the
coupling function in units of $t$. The right plot shows the flow of the AF
susceptibility (dashed line), $d$-wave susceptibility (solid line) and FM
susceptibility (dotted line) as a function of the temperature $T$.}
\label{vp100}
\end{figure}

If we increase the absolute value of $t'$ and adjust the chemical potential
such that the FS remains at the van Hove points, the characteristic
temperature $T_c$ for the flow to strong coupling drops continuously 
and for $t'< -0.2t$, the $d$-wave
susceptibility takes over as the leading susceptibility. In the flow of the
couplings one clearly observes the dominant $d_{x^2-y^2}$-symmetry of the pair
scattering, see the diagonal features in the middle plot of Fig. \ref{vp100}. 
For $t' \approx -0.25t$  and a band filling slightly larger than
the van Hove filling we again find a regime where the flow of $d$-wave and AF
processes is strongly coupled similar as the saddle point regime studied in
Ref. \onlinecite{honerkamp}. 

\begin{figure} 
\begin{center} 
\includegraphics[width=.6\textwidth]{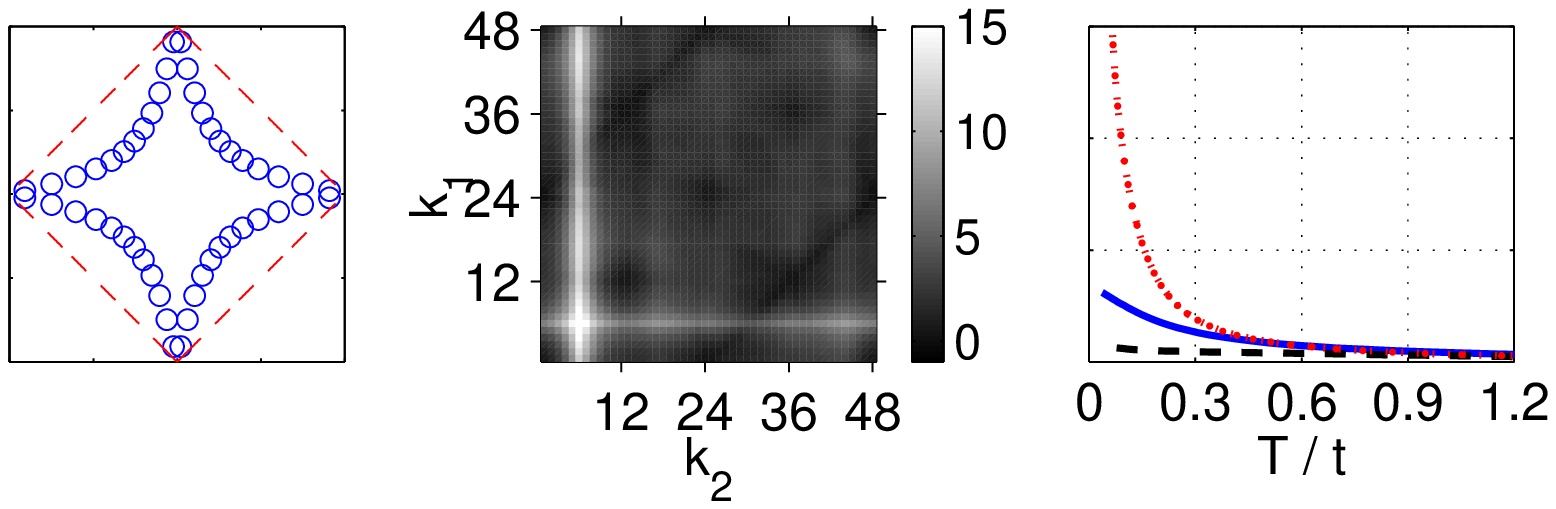}
\end{center}
\caption{Flow to strong coupling for $t'=-0.45t$ and $\mu=-1.8t$. The left
plot shows the 48 points on the Fermi surface (points 1, 12, 13, 24 etc. are
closest to the saddle points). The middle plot shows the
coupling function $V_T(k_1,k_2,k_3)$ at low $T$ as a function of the two
incoming wavevectors labeled by $k_1$ and $k_2$ moving around the FS. $k_3$
is fixed at point 6 in the Brillouin diagonal. The colorbar denotes the value of the
coupling function in units of $t$. The right plot shows the flow of the AF
susceptibility (dashed line), $d$-wave susceptibility (solid line) and FM
susceptibility (dotted line) as a function of the temperature $T$.}
\label{vp180}
\end{figure}

Here we focus on the flow at the van Hove filling when we increase the
absolute value of $t'$. The characteristic temperature for the flow to strong
coupling drops rapidly for $t' \le
-0.3t$, while for $t' \ge -0.33t$ it rises again. For these values the flow to 
strong coupling is dominated by processes with small momentum transfer (middle 
plot in Fig. \ref{vp180}) and the
FM susceptibility $\chi_s (\vec{q} = 0)$ is by far the most divergent
susceptibility at low temperatures (right plot in Fig. \ref{vp180}).
 The overall behavior strongly suggests a
critical value for $t'$ around $t'_c \approx -0.33t$ 
where the ground state changes from
$d$-wave singlet superconducting to ferromagnetic. We note that while our method
allows, within the approximations made, to detect the instability of the Fermi
liquid state against ferromagnetic fluctuations it does not give any
information on the degree of the polarization of the ordered ground state.


Further we stress the fundamental difference of the qualitative change
in the flow between the $d$-wave to the FM regime to the crossover from the AF 
to the $d$-wave regime: in the latter 
case the transition is continuous and takes place at
relatively high scales. As emphasized in Refs. \onlinecite{honerkamp} and
\onlinecite{honerkamp2}, for the FS close to the saddle points and away from half-filling, AF and $d$-wave 
tendencies do not compete but reinforce each other on the one-loop level,
therefore one finds a gradual change in the character of the flow to strong
coupling when $t'$ is varied. There are strong indications\cite{honerkamp} 
that the strong coupling state for a certain parameter range is not
simply a symmetry--broken phase.
In contrast with that, the transition from the $d_{x^2-y^2}$-wave regime to the
FM regime at larger absolute values of $t'$ is very distinct and - as
suggested by our one-loop analysis -  may be a quantum critical point of two 
mutually excluding tendencies. Due to the competition
between singlet superconducting and ferromagnetic tendencies, the characteristic
temperature for the flow to strong coupling becomes suppressed to smallest values around $t'_c$. The precise properties of the quantum critical point cannot be 
analyzed further for the time being because the
numerical integration of the RG flow becomes rather time-consuming 
for these parameters.

The rivalry between singlet superconducting and FM tendencies
is already foreshadowed on the $d$-wave side at $t' \approx -0.3t$. There 
the temperature-flow scheme yields much lower characteristic temperatures than the
momentum-shell scheme used in Ref. \onlinecite{honerkamp} which is rather insensitive
against the FM tendencies arising from the van Hove points for the reasons
mentioned in the introduction. 

\begin{figure} 
\begin{center} 
\includegraphics[width=.65\textwidth]{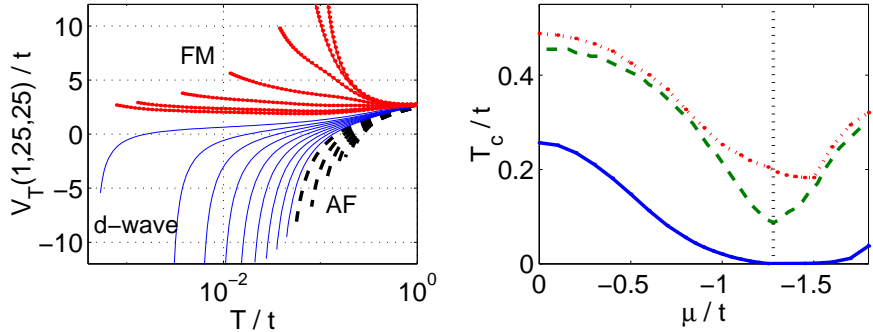}
\end{center}
\caption{Left: Flow of the forward scattering coupling function $V_T (1,25,25)$ for
the saddle point filling and different values of $t'$. The dashed (solid) lines
correspond to the AF (d-wave) regime, where $V_T (1,25,25)$ decreases when
the temperature $T$ is lowered, while the dotted lines correspond to the FM regime,
where $V_T (1,25,25)$ increases with decreasing $T$.
Right:  Characteristic temperature $T_c$ for the flow to strong coupling 
versus $t'$ at the van Hove
filling $\mu = 4t'$. The lower curves (same as in Fig. \ref{tctpvh}) are
obtained with the full one-loop RG, while the upper dotted line result from the RG 
where the particle-particle channel is left out.  Right plot: Comparison of the flow of the FM susceptibility
$\chi_s(0)$ with and without particle-particle diagrams. For both cases, the RG
equations were integrated until the couplings reach $V_{T, \mathrm{max.}}
\approx 18t$. The dashed line denotes the critical temperature from the
generalized Stoner criterion. The vertical line at $t'=-0.32t$ separates the
FM regime from the AF in the Stoner calculation and RG without particle-particle
diagrams. }
\label{v12424}
\end{figure}

The transition from singlet pairing to the 
ferromagnetic regime can also be seen in the flow of certain coupling
functions which couple into both channels. For example let us take the
component $V_T(1,25,25)$
for a $N=48$ system (see Fig. \ref{v12424}). The incoming wavevectors 
$\vec{k}_{F,1}$ and $\vec{k}_{F,25}$ add up to zero, therefore the coupling
corresponds to a Cooper process where a pair of electrons is scattered onto
itself. 
Since wavevector $\vec{k}_{F,1}$ is close to the saddle points it should
rapidly diverge to $-\infty$ 
if the flow is towards a $d$-wave superconducting ground state. On
the other hand  $V_T(1,25,25)$ couples into the FM channel as well, as the
momentum transfer between the second incoming and the first outgoing
wavevector is the same (in the calculation we use a very small momentum
transfer). In Landau-Fermi liquid  language $V_T(1,25,25)$ corresponds to the
process $a^a (\vec{k}_{F,1}, -\vec{k}_{F,1} )$, which should become large and
positive if the flow signals a FM ground state. By plotting the flow of
$V_T(1,25,25)$ for the saddle point filling and different $t'$-values, as
shown in Fig. \ref{v12424} we can
nicely observe the qualitative change in the flow when $|t'|$ becomes larger
than the critical value $t'_c \approx -0.33t$. We note that $V_T(1,25,25)$ does
not belong to the dominant scattering processes in the FM regime. The latter
are given by forward scattering processes $a^a (\vec{k}_F, \vec{k}'_F)$ with
$\vec{k}_F$ and $\vec{k}'_F$
closer to the BZ diagonals (see Fig. \ref{vp180}) and grow more strongly
towards $T_c$.

One can also investigate the contributions of different one-loop diagrams to
the RG flow. Here we concentrate on the influence of the particle-particle channel
on the flow. In the left plot of Fig. \ref{v12424} we compare the characteristic temperatures for the
full one-loop RG with the results when the particle-particle channel is left 
out. For all values of $t'$, the  $T_c$ is much higher, which shows clearly
the importance of the particle-particle channel for a screening of the repulsive
interaction. Without the particle-particle diagrams and close to $t'=0$ and
 $t'=-0.5t$ the critical temperatures from the RG and from the generalized Stoner-criterion\cite{fn3} are very similar.
Again the FM susceptibility becomes dominant around $t'_c \approx
-0.33t$, but without the particle-particle diagrams generating strong
superconducting fluctuations, the characteristic temperature for the flow to
strong coupling is only weakly suppressed for these parameters.

\section{The flow away from the van Hove filling}
Next we investigate the flow to strong coupling for band fillings away from the 
van Hove filling. The flow arising for smaller absolute values of the
next-nearest neighbor hopping $t'$ is very similar to the flow found with the
momentum-shell techniques which has been extensively discussed in
Refs. \cite{zanchi,halboth,honerkamp,honerkamp2}. Therefore we now focus on 
the case $t'<-0.33t$  where we find a flow dominated by ferromagnetic
fluctuations at the van Hove filling. 
Changing the particle density moves the van Hove
singularities away from the Fermi surface and one expects a reduction of the
characteristic temperature for the flow to strong coupling accompanied with a
decrease of the ferromagnetic tendencies. Furthermore, by similar arguments as 
for the 
interplay between antiferromagnetic and $d$-wave superconducting fluctuations
when the FS nesting is reduced, one may presume that some kind of triplet
superconducting channel might diverge at lower temperatures where the
ferromagnetic tendencies get cut off.
This expectation is confirmed by the results of the temperature-flow RG
scheme. 

The Cooper pair scattering $V_T(\vec{k},-\vec{k},\vec{k}')$ transforms
according to one of the 5 irreducible representations of the point group of
the 2D square lattice. Among those there is the one-dimensional representation with
$d_{x^2-y^2}$-symmetry important  for smaller $|t'|$ and the only
two-dimensional representation transforming like $p_x$ and $p_y$. In a
solution of the BCS gap equation, components belonging to different
representations do not mix, while e.g. within the $p$-representation
components $p_x \propto \cos \theta$ and $p_y \propto \sin \theta$ and
the higher harmonics $\propto \cos 3 \theta$ and $\propto 
\sin 3\theta$ (also called $f$-wave) will occur together.

\begin{figure} 
\begin{center} 
\includegraphics[width=.6\textwidth]{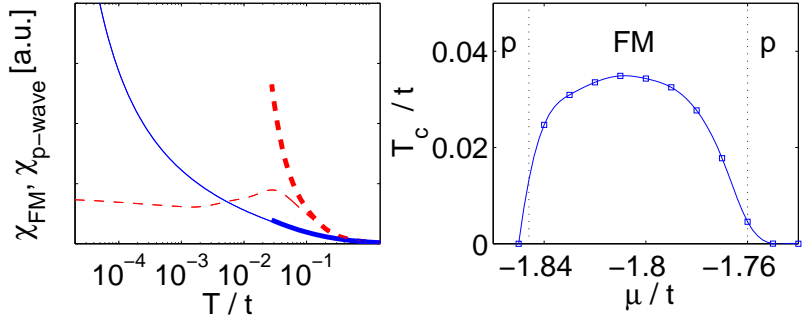}
\end{center}
\caption{Left plot:
Flow of ferromagnetic (solid lines), $p_x$-wave (dashed lines)
 for $t'=-0.45t$ at $\mu=-1.78t$ (thick lines) and $\mu=-1.74t$ (thin
lines). Right plot: Critical
temperature $T_c$ where the largest couplings reach $18t$ as function of the
chemical potential $\mu$. The dotted vertical lines separate the parameter
region where the FM or the $p$-wave susceptibility grow most strongly.}
\label{tp45pw}
\end{figure}

\begin{figure} 
\begin{center} 
\includegraphics[width=.65\textwidth]{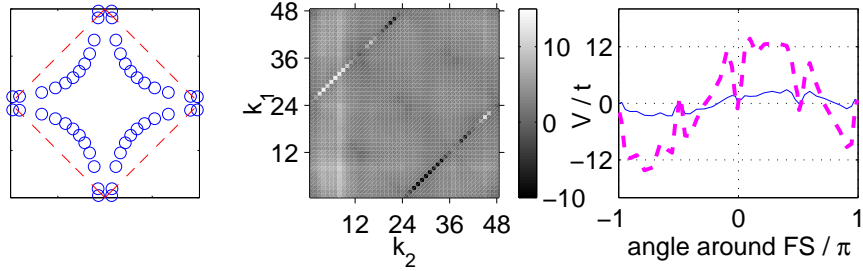}
\end{center}
\caption{Left panel: Fermi surface for $\mu =-1.71t$ and $t'=-0.45t$. Middle
panel: snapshot of the coupling function $V(k_1,k_2,k_3)$ with $k_3=6$ fixed in 
the Brillouin zone diagonal. Right panel: Pair scattering
$V(\vec{k},-\vec{k},\vec{k}_3)$ with $\vec{k}$ varying around the Fermi
surface at $T=1.1\cdot 10^{-6} t$.. For the solid line $\vec{k}_3$ is fixed at point 1 close to the
saddle point $(-\pi,0)$ while for the dashed line, $\vec{k}_3 \approx
(-1.06,-0.86)$ is at point 6.}
\label{pwpscat}
\end{figure}

As the RG calculations down to low
temperatures are rather time-consuming we limit ourselves to the exemplary
case $t'=-0.45t$ and leave a full investigation of the
parameter space to future work.
As shown in the right plot of
Fig. \ref{tp45pw} for $t'=-0.45t$, the critical  temperature
drops by several orders of magnitude when we increase the particle density per 
site from the VH value $\langle n \rangle \approx 0.47$ at $\mu=-1.8t$
to $\langle n \rangle \approx 0.58$ at $\mu=-1.7t$. Further upon moving away from the VH filling, the growth of the
FM susceptibility gets cut off and the $p$-wave 
triplet superconducting
susceptibilities with symmetry $p_x\propto \cos \theta$ or $p_y\propto \sin
\theta$ diverge at low
temperature. Higher order harmonics $\propto \cos 3 \theta$ and $\sin
3\theta$ diverge in a weaker fashion.
The pair scattering at low temperatures in
this parameter range is shown in Fig.\ref{pwpscat}. 
One nicely observes that the pair scattering
involving particles close to the saddle points is suppressed as the odd-parity 
nature of the $p$-wave pairing requires an opposite sign  of the pair
scattering $V_T (\vec{k}, -\vec{k}, \vec{k}')$ e.g. for $\vec{k}' \approx (0,
\pi)$ and $\vec{k}' \approx (0, -\pi)$.
Comparing the temperature derivatives of ferromagnetic and
superconducting susceptibilities when the leading couplings become larger than
$12t$, the transition from the ferromagnetic to the $p$-wave superconducting
regime occurs at a particle density of $\langle n \rangle \approx 0.55$ per
site. Again, very similar to the interplay between antiferromagnetic and $d_{x^2-y^2}$
superconducting fluctuations for small $|t'|$, 
on the one-loop level there is a smooth evolution of the flow to strong coupling from 
the strongly FM dominated to the predominantly $p$-wave dominated instability. This
suggests the presence of a transition between the two types of
ordered states as a function of the band filling 
provided that symmetry-breaking at finite temperatures becomes 
possible in a three-dimensional environment. From our analysis the transition
appears to be first order, but additional interaction effects beyond our
one-loop calculation could change the behavior.


It is plausible to assume that the superconducting state for
$t'=-0.45t$ will be given by
a nodeless superposition of the two components with
symmetries $p_x$ and $p_y$, which 
maximizes the condensation energy\cite{rice}. 
The direction of the Cooper pair spin or the $\vec{d} (\vec{k})$-vector 
parameterizing the odd-parity gap function 
remains indeterminate in the absence of spin-orbit coupling. Therefore six
different pairing symmetries $\vec{d} (\vec{k}) \propto\vec{x} k_x \pm \vec{y}
k_y$, $\propto\vec{x} k_y \pm \vec{y}
k_x$ and $\propto\vec{z} (k_x \pm i k_y)$ yield the same condensation energy
and additional interactions like spin orbit coupling or the spin fluctuation
feedback mechanism are needed to select 
one of these representations\cite{ng}.
In any case the superconducting gap function for given $s$ and $s'$  will read 
\begin{equation} \Delta_{ss'} (\vec{k}) = \Delta_0 (\vec{k}) \, (k_x \pm i k_y) \, , 
\label{kxiky}
\end{equation}
where the real prefactor $\Delta_0 (\vec{k})$ is unchanged under symmetry
transformations of the lattice and takes care of the strong
anisotropy within symmetry-related Fermi surface parts. 
 
\begin{figure} 
\begin{center} 
\includegraphics[width=.6\textwidth]{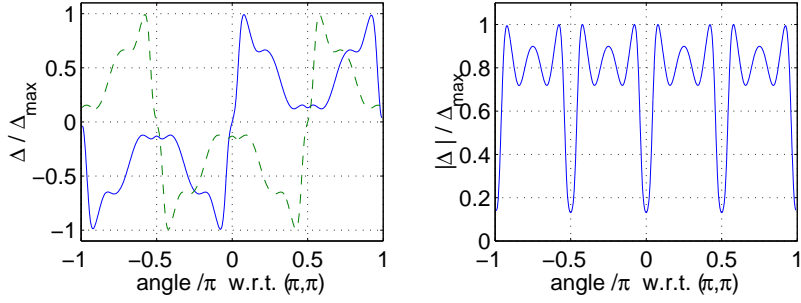}
\end{center}
\caption{Left panel: Angular variation of real (solid line) and imaginary
(dashed line) part of the gap function obtained by the method explained in the
text. Right panel: Angular variation of the gap magnitude. For these data,
$\mu=-1.71t $ and $t'=-0.45t$.}
\label{gapeq}
\end{figure}

Our present method does not allow a straightforward calculation of the gap
function in the superconducting state. Nevertheless we can use the Cooper
pair scattering obtained from the RG flow as effective pair potential in a
mean-field treatment of the superconducting state. Since the Cooper scattering 
diverges in our approach, this will not give any
precise information on the overall magnitude of the energy gap, but it allows
us to study the symmetry and the angular variation of the gap function. In
Fig. \ref{gapeq} we show the solution of the BCS gap equation with symmetry
given by Eq. (\ref{kxiky}) and using the rescaled 
Cooper pair scattering extracted from the RG in the case 
$\mu=-1.71t$ and $t'=-0.45t$. 
For these data the angle is measured with respect to the $(\pi,\pi)$-
point. Clearly the dips in the pair scattering at the saddle points
mentioned above and observable in Fig. \ref{pwpscat} lead to minima of the
absolute gap magnitude on these FS parts. Thus the particular location of the
FS close to the BZ boundary results
in a highly anisotropic superconducting energy gap.

\section{Comparison with other approaches and conclusions}
We have described a modified RG scheme for interacting fermion systems 
which uses the temperature as explicit
scaling parameter.  In comparison with the momentum-shell schemes
\cite{msbook,zanchi,halboth,salmhofer} the new approach 
has the advantage that it allows one to treat
small-$\vec{q}$ singularities in the particle-hole channel, in our case corresponding
to ferromagnetic tendencies, on the same basis as large-$\vec{q}$
(here antiferromagnetic) fluctuations or superconducting correlations. 

At first sight the successive lowering of the temperature appears to be closer
to the physical intuition than the reduction of an momentum-shell cutoff. 
On the other hand it should be mentioned that in general the initial conditions
for the vertex function of  the temperature-flow RG at the starting
temperature do not exactly equal the vertex functions (to some order) of the
microscopic Hamiltonian. The latter fact might complicate the comparison to other
techniques.  

Using the temperature-flow RG scheme, 
we have analyzed the flow to strong coupling for the two-dimensional $t$-$t'$
repulsive Hubbard model on the square lattice. 
The results obtained agree qualitatively with a various other calculations. 

The flow to strong coupling in the  parameter region
of small next nearest neighbor hopping $t'$ shows strong similarities to the RG flow 
known from the momentum-shell techniques\cite{zanchi,halboth,honerkamp}. 
Again AF and $d$-wave tendencies are 
coupled to a large extent and the FS curvature determines which of the two
types of fluctuations is strongest: when we move away from perfect nesting,
the AF processes are cut off at low temperatures and the $d_{x^2-y^2}$-wave 
Cooper scattering processes can diverge.  

For larger absolute values of $t'$ beyond a critical $t'_c \approx -0.33t$ 
we find that the flow to strong coupling is dominated by ferromagnetic fluctuations.
This regime has not been found with the momentum-shell schemes used
recently\cite{zanchi,halboth,honerkamp} for the 
reasons explained in the introduction. 

There are several other approaches which predict ferromagnetic
ground states for larger $|t'|$. Lin and Hirsch\cite{linhirsch} analyzed the
Hartree-Fock phase diagram for the $t$-$t'$ Hubbard model and found the
ferromagnetic state to be stable against the antiferromagnetic state for $t' < 
-0.324t$. They do not find a $d$-wave phase for $t'>-0.324t$ because 
the bare repulsive Hubbard interaction does not contain any attractive
component in the pair scattering channel.
The critical $t'$-value from Hartree-Fock is practically equivalent to the
critical $t'_c \approx -0.33t$ from the RG treatment. 
On the other hand in Hartree-Fock one
would find a first order transition at $t'_c$, while in our
treatment the characteristic temperature gets suppressed to zero at the quantum
critical point. Irkhin et al.\cite{irkhin} applied a parquet scheme using the
same set of one-loop diagrams as the RG described above. With a simplified
dispersion they found indications for 
ferromagnetism close to $t'=-0.5t$ and emphasized the strong
deviations  of the critical temperatures from the simple Stoner calculation.
Hlubina et al. \cite{hlubina1,hlubina2} analyzed the question of
ferromagnetism by applying quantum Montecarlo and T-matrix approximation
(TMA). The latter method gave a ferromagnetic state beyond a critical $t' \approx
-0.43t$. The Kanamori-screening in the particle-particle channel of the TMA 
 is included 
in our temperature-flow scheme, which gives a larger window for a
ferromagnetic ground state. This suggests that the straightforward TMA might
overestimate the screening effects in the given example. We repeat however
that defining
the initial condition for two-point and four-point vertex functions in 
the temperature-flow  RG is not fully equivalent to
using a specific microscopic Hamiltonian in the perturbative scheme. 
Arachea\cite{arachea} found spin-polarized ground states in exact
diagonalization studies of a $4\times 4$ $t$-$t'$ cluster for $t'=-0.4t$ and
$t'=-0.6t$ at densities 0.5 and 0.375.

For several classes of models\cite{fazekas,tasaki,mielke,vollhardt} large-spin ground
states can be proved exactly. Among those, the so-called flat band
ferromagnetism appears to have the closest relation to the case of the $t$-$t'$
Hubbard model. 
The typical feature of these models is a large density of states at the bottom of the band which gets pushed further
below the Fermi energy by polarizing the system. In the 2D $t$-$t'$ Hubbard
model for $t' \to -0.5t$
 the density of states is peaked at the bottom of the
band. Nonetheless from the perspective of our RG calculation 
the large density of states at the FS appears to be essential for the FM 
tendencies in the $t$-$t'$ Hubbard model, as the latter
are strongly reduced when one raises the Fermi level above the van Hove
energy. Moreover, if we cut out the bottom of the band by hand and discard all
modes with $\epsilon_{\vec{k}}<-0.1t$, the
ferromagnetic tendencies do increase rather than decrease, presumably due to the
reduced screening of the onsite repulsion.
Thus the peak at the bottom of the band is not essential for our finding
of ferromagnetic tendencies.
The sensitivity of the ferromagnetic regime to the location of the FS
points to another effect which is not included in the present calculation.
 It is probable that the inclusion
of selfenergy effects in the flow will shift the Fermi surface and alter the
low energy density of states such that the actual ferromagnetic regime might
be changed. A calculation of the selfenergy at the FS from the
flow of the interactions shows however that the real part of the selfenergy
becomes comparable to the temperature only when the couplings have grown
larger than the bandwidth to $V_{T,\mathrm{max.}} \approx 12-13t$.

In analogy with the interplay between AF and $d$-wave fluctuations close to
perfect nesting at smaller $t'$,  there exists a parameter region for 
larger $|t'|$ where the FM tendencies get cut off
and the flow to strong coupling is dominated by $p_x$- and $p_y$-wave
superconducting correlations. The energy gap suggested by the RG flow of the Cooper
pair scattering is highly anisotropic and exhibits minima on the FS parts close to 
the saddle points. Similar scenarios have already been proposed by
several authors\cite{hlubina1,hlubina2,guinea}, our approach allows a more
systematic analysis of the crossover from the FM to $p$-wave regime and the
possible order parameter symmetries.
We note that the superconducting gap functions
suggested by the RG flow $\propto p_x+ ip_y$ are among the candidates
under debate for the superconducting state of Sr$_2$RuO$_4$
\cite{physicstoday}. 
In this quasi-two-dimensional system one of the three Fermi surfaces has a
similar shape as the cases studied above. 
It should be interesting to investigate the RG flow and the superconducting
properties suggested by the resulting pair scattering in a more realistic 
three-band model.
 
Another future application of the modified RG scheme could be to investigate
the interplay of screening of long-range forces with other tendencies. 
Further, other potential 
instabilities in forward-scattering channel could be analyzed, e.g.  Labbe-Friedel or
Pomeranchuk instabilities with spontaneous deformations of the FS, as proposed
by Halboth and Metzner\cite{halboth}. \\[.2cm]

We are grateful to T.M. Rice, M. Sigrist, R. Hlubina, P. Monthoux, M.E. Zhitomirsky and
D. Vollhardt for valuable
discussions. C.H. acknowledges financial support through the Swiss National Science Foundation.


\begin{thebibliography}{99}
\bibitem{fn2} A flow where the particle number is fixed instead of $\mu$
gives qualitatively similar results. The same holds for smaller initial
interactions, e.g. $U=2.5t$.
\end{thebibliography}
\end{document}